\documentclass{PoS}

\newcommand{\beqn}{\begin{eqnarray}}
\newcommand{\eeqn}{\end{eqnarray}}
\newcommand{\be}{\begin{equation}}
\newcommand{\ee}{\end{equation}}

\newcommand{\ba}{\begin{array}{c}}
\newcommand{\bat}{\begin{array}{cc}}
\newcommand{\ea}{\end{array}}

\newcommand{\bi}{\begin{itemize}}
\newcommand{\ei}{\end{itemize}}

\newcommand{\ket}{\,\rangle}
\newcommand{\bra}{\langle \,}

\newcommand{\Frac}[2]{\frac{\displaystyle #1}{\displaystyle #2}}

\newcommand{\Int}{\displaystyle{\int}}

\title{One-loop calculation of the oblique S and T parameters within strongly-coupled scenarios with a light Higgs-like boson }

\ShortTitle{One-loop calculation of the oblique S and T parameters within strongly-coupled scenarios}

\author{Ignasi Rosell $^{ab}$\thanks{We wish to thank the organizers of CD12 for the pleasant congress and A.~Pich and J.J.Sanz-Cillero for their comments. This work has been supported in part by the Spanish Government and ERDF funds from the EU Commission [grants FPA2007-60323, FPA2011-23778 and CSD2007-00042 (CPAN)] and the Universidad CEU Cardenal Herrera [grant PRCEU-UCH35/11]. }\\
\llap{$^a$} Departamento de Ciencias F\'\i sicas, Matem\'aticas y de la Computaci\'on,  \\ Escuela Superior de Ense\~nanzas T\'ecnicas ESET, \\ Universidad CEU Cardenal Herrera, \\ c/ Sant Bartomeu 55, E-46115 Alfara del Patriarca, Val\`encia, Spain \\
\llap{$^b$} IFIC, Universitat de Val\`encia -- CSIC,\\ Apartat de Correus 22085, E-46071 Val\`encia, Spain\\
        
        \email{rosell@uch.ceu.es}}

\abstract{We present a one-loop calculation of the oblique $S$ and $T$ parameters within
strongly-coupled models of electroweak symmetry breaking with a light Higgs-like boson.
We use a general  effective Lagrangian, implementing the chiral symmetry breaking $SU(2)_L\otimes SU(2)_R\to SU(2)_{L+R}$ with Goldstones, gauge bosons,
the Higgs-like scalar and one multiplet of vector and axial-vector massive resonance states.
The estimation is based on the short-distance constraints and a dispersive approach.
The experimentally allowed range forces the vector and axial-vector states to be heavy, with masses
above the TeV scale, and suggests that the Higgs-like scalar should have a $WW$ coupling close to the Standard Model one.}

\FullConference{The 7th International Workshop on Chiral Dynamics,\\
		August 6 -10, 2012\\
		Jefferson Lab, Newport News, Virginia, USA}

\begin{document}

\section{Introduction}

A new Higgs-like boson around $126\,$GeV has just been discovered at the LHC~\cite{LHC}.
Although its properties are not well measured yet, it complies with the expected behaviour
and therefore it is a very compelling candidate to be the Standard Model (SM) Higgs.
An obvious question to address is to which extent alternative scenarios of
Electroweak Symmetry Breaking (EWSB) can be already discarded or strongly constrained.
In particular, what are the implications for strongly-coupled models where the
electroweak symmetry is broken dynamically?

The existing phenomenological tests have confirmed the
$SU(2)_L\otimes SU(2)_R\rightarrow SU(2)_{L+R}$
pattern of symmetry breaking, giving rise to three Goldstone bosons which, in the
unitary gauge, become the longitudinal polarizations of the gauge bosons.
When the $U(1)_Y$ coupling $g'$ is neglected, the electroweak Goldstone dynamics
is described at low energies by the same Lagrangian as
the QCD pions, replacing the pion decay constant by the
EWSB scale $v=(\sqrt{2}G_F)^{-1/2} = 246\,$GeV~\cite{AB:80}.
In most strongly-coupled scenarios the symmetry is nonlinearly realized and one expects the appearance of massive resonances generated by the non-perturbative interaction.

The dynamics of Goldstones and massive resonance states can be analyzed in a generic way by using
an effective Lagrangian, based on symmetry considerations.
The theoretical framework is completely analogous to the Resonance Chiral Theory description
of QCD at GeV energies~\cite{RChT}.
Using these techniques, we have investigated in Ref.~\cite{paper2}, and as an update of Ref.~\cite{paper},  the oblique $S$ and $T$ parameters~\cite{Peskin:92}, characterizing the
electroweak boson self-energies, within strongly-coupled models that incorporate a light Higgs-like boson.
Adopting a dispersive approach and imposing a proper high-energy behaviour, it has been shown there that it is possible
to calculate $S$ and $T$ at the next-to-leading order, {\it i.e.}, at one-loop.
We have found that in most strongly-coupled scenarios of EWSB a high resonance mass scale is required, above $1\,$TeV,
to satisfy the stringent experimental limits.
Previous one-loop analyses can be found in Refs.~\cite{other}.

\section{Theoretical Framework}

We have considered a low-energy effective theory containing the SM gauge bosons coupled
to the electroweak Goldstones, one light scalar state $S_1$ with mass $m_{S_1} = 126$~GeV
and the lightest vector and axial-vector resonance multiplets $V_{\mu\nu}$ and $A_{\mu\nu}$.
We have only assumed the SM pattern of EWSB, {\it i.e.} the theory
is symmetric under $SU(2)_L\otimes SU(2)_R$ and becomes spontaneously broken to the diagonal
subgroup $SU(2)_{L+R}$.
$S_1$ is taken to be singlet under $SU(2)_{L+R}$, while $V_{\mu\nu}$ and $A_{\mu\nu}$ are triplets. 
To build the Lagrangian we have only considered operators with the lowest number of derivatives,
as higher-derivative terms are either proportional to the equations of motion or tend to violate the expected short-distance behaviour~\cite{RChT}.
We have needed the interactions~\cite{paper2}
\begin{equation}\label{eq:Lagrangian}
\mathcal{L}\!\!=\!\!
\frac{v^2}{4}\!\bra \! u_\mu \! u^\mu \!\!\ket \!\!\left(\!\!\! 1\! +\! \frac{2\omega}{v} S_1\!\!\!\right)\!\!
 +\! \frac{F_A}{2\sqrt{2}} \!\bra \!A_{\mu\nu} \!f^{\mu\nu}_- \!\ket
\!+\! \frac{F_V}{2\sqrt{2}} \!\bra \!V_{\mu\nu} \!f^{\mu\nu}_+ \!\ket
\!+\! \frac{i G_V}{2\sqrt{2}} \!\bra \! V_{\mu\nu} [u^\mu\! , u^\nu] \!\ket
\!+\! \sqrt{2} \lambda_1^{SA}  \partial_\mu S_1  \!\bra \! A^{\mu \nu}\! u_\nu \!\ket \!,
\end{equation}
plus the standard gauge boson and resonance kinetic terms.
We have followed the notation from Ref.~\cite{paper}.
The first term in (\ref{eq:Lagrangian}) gives the Goldstone Lagrangian, present in the SM, plus the
scalar-Goldstone interactions.
For $\omega=1$ one recovers the $S_1\to\pi\pi$ vertex of the SM. 

The oblique parameter $S$ receives tree-level contributions from vector and axial-vector exchanges \cite{Peskin:92}, while
$T$ is identically zero at lowest-order (LO):
\begin{equation}
S_{\mathrm{LO}} = 4\pi \left( \frac{F_V^2}{M_V^2}\! -\! \frac{F_A^2}{M_A^2} \right)  \,,
\qquad\quad
T_{\mathrm{LO}}=0 \,.
\label{eq:LO}
\end{equation}
To compute the one-loop contributions we have used the dispersive representation of $S$ introduced
by Peskin and Takeuchi~\cite{Peskin:92},
whose convergence requires a vanishing spectral function at short distances:
\begin{equation}
S\, =\, \Frac{16 \pi}{g^2\tan\theta_W}\,
\Int_0^\infty \, \Frac{{\rm dt}}{t} \, [\, \rho_S(t)\, - \, \rho_S(t)^{\rm SM} \, ]\, , \label{Sintegral}
\end{equation}
with $\rho_S(t)\,\,$ the spectral function of the $W^3B$ correlator~\cite{paper2,paper,Peskin:92}.
We have worked at lowest order in $g$ and $g'$ and only the lightest cuts have been considered, {\it i.e.} two Goldstones or one Goldstone plus one scalar
resonance. $V\pi$ and $A\pi$ contributions were shown to be suppressed in Ref.~\cite{paper}.

The calculation of $T$  is simplified by noticing that, up to corrections of $\mathcal{O}(m_W^2/M_R^2)$, $T=Z^{(+)}/Z^{(0)}-1$,
being $Z^{(+)}$ and $Z^{(0)}$ the wave-function renormalization constants of the charged
and neutral Goldstone bosons computed in the Landau gauge~\cite{Barbieri:1992dq}.
A further simplification occurs by setting to zero $g$, which does not break the custodial symmetry, so
only the $B$-boson exchange produces an effect in $T$.
This approximation captures the lowest order contribution to $T$   in its expansion
in powers of   $g$ and $g'$.
Again only the lowest two-particle
cuts have been considered, {\it i.e.} the $B$ boson plus one Goldstone or one scalar resonance.

%
Requiring the
$W^3 B$ correlator to vanish at high energies leads to a good convergence of the Goldstone
self-energies, at least for the cuts we have considered.
Then, their difference obeys an unsubtracted
dispersion relation, which enables us to compute $T$ through the dispersive integral~\cite{paper2},
\begin{eqnarray}
T &=& \Frac{4 \pi}{g'^2 \cos^2\theta_W}\, \Int_0^\infty \,\Frac{{\rm dt}}{t^2} \,
[\, \rho_T(t) \, -\, \rho_T(t)^{\rm SM} \,] \, , \label{Tintegral}
\end{eqnarray}
with $\rho_T(t)\,\,$
 the spectral function of the difference of the neutral and charged Goldstone self-energies.

\section{The calculation}

The spectral functions of Eqs.~(\ref{Sintegral}) and (\ref{Tintegral}) read:
\begin{eqnarray}
\rho_S(s)|_{\pi\pi} &=& \Frac{g^2\tan\theta_w}{192\pi^2}\, \bigg(1+\kappa_V \Frac{s}{M_V^2-s}\bigg)^2
\,\theta(s)\, , \label{Spipi} \\
\rho_S(s)|_{S\pi} &=& -\, \Frac{g^2\tan\theta_w}{192\pi^2}\, \, \omega^2 
\bigg(1+\kappa_A \Frac{s}{M_A^2-s}\bigg)^2
\,\bigg(1-\Frac{m_{S_1}}{s}\bigg)^3\, \theta(s-m_{S_1}^2)\, ,  \label{SSpi}\\
\rho_S(s)|_{SM} &=& \Frac{g^2 \tan\theta_W }{192\pi^2}\, \bigg[ \theta(s)  \, -\, \bigg(1-\Frac{m_H}{s}\bigg)^3\theta(s-m_H^2)\bigg]\, , \\
\rho_T(s)|_{B \pi} &=& - \Frac{g'^2s}{64\pi^2}\bigg[ \left(3-2\,\hat{s}\, \kappa_V \right)\theta(s)+\kappa_V\left(1-\frac{1}{\hat{s}}\right)^2 \left( 3\kappa_V+2\,\hat{s} -2\right) \theta(\hat{s}-1) \bigg] \,,\\
\rho_T(s)|_{B S} &=& \Frac{g'^2\omega ^2 s}{64\pi^2} \bigg[ \bigg(3\bigg(1-\frac{m_{S_1}^4}{s^2} \bigg)-2\widetilde{s} \kappa_A\bigg(1-\frac{m_{S_1}^2}{s} \bigg)^3\bigg)\theta(s-m_{S_1}^2) 
\nonumber \\ && \qquad \qquad \qquad \qquad \qquad 
+ \kappa_A \left(1-\frac{1}{\widetilde{s}}\right)^2 \left( 3\kappa_A+2\,\widetilde{s} -2\right)\theta( \widetilde{s} -1) \bigg] \,, \label{hola}\\
\rho_T(s)|_{SM} &=& \Frac{3{g'}^2s}{64\pi^2}  \bigg[-\theta(s)+\left(1-\frac{m_H^4}{s^2} \right)\theta (s-m_H^2)\bigg] \,,
\end{eqnarray}
being $\kappa_V=F_VG_V/v^2$, $\kappa_A=F_A \lambda^{SA}_1/(\omega v)$, $\hat{s}=s/M_V^2$ and $\widetilde{s}=s/M_A^2$. Terms of $\mathcal{O}(m_{S_1}^2/M_{V,A}^2)$ have been neglected in Eq.~(\ref{hola}). 

Fixing $m_{S_1}=126$~GeV, one has 7 undetermined parameters: $M_V$, $M_A$, $F_V$, $G_V$, $F_A$, $\omega$ and $\lambda_1^{SA}$.
The number of unknown couplings can be reduced using short-distance information~\cite{paper2}:
\begin{enumerate}
\item {\bf  Vector form factor}. The two-Goldstone matrix element of the vector current defines the vector form factor (VFF). Imposing that it vanishes at $s\rightarrow \infty$, one finds that $F_V G_V  = v^2$~\cite{RChT}.
\item {\bf Weinberg sum rules at leading order}. Assuming the two Weinberg sum rules (WSRs)~\cite{WSR} at leading order one gets
\begin{equation}
 F_{V}^2 \,-\, F_{A}^2 \, =\, v^2 \, , \qquad \qquad F_{V}^2 \, M_{V}^2\, -\, F_{A}^2 \, M_{A}^2  \,=\, 0 \,  .
\end{equation}
This implies $M_A > M_V$ and determines $F_V$ and $F_A$ in terms of the resonance masses. 
 Note that the second WSR is questionable in some scenarios.
\item {\bf Weinberg sum rules at one loop}. At next-to-leading order the computed spectral functions of Eqs.~(\ref{Spipi}) and (\ref{SSpi}) should behave also as dictated by this pattern. Once the constraint coming from the VFF has been used, the first and the second WSRs provide respectively~\cite{paper2} 
\begin{equation}
F_A \lambda^{SA}_1 \,=\, \omega v \,, \qquad \qquad \quad \omega \,=\,  M_V^2/M_A^2 \,.
\end{equation}
After imposing the short-distance conditions on the spectral function, one has to apply the same constraints to the real part of the correlator, reaching the next-to-leading extension of the first and second Weinberg sum rules~\cite{paper}, respectively,
\begin{eqnarray}
F_{V}^{r\,2} \, -\, F_{A}^{r\,2}\; = \; v^2\, (1\,+\,\delta_{_{\rm NLO}}^{(1)}) \, ,\quad\;
\qquad 
F_{V}^{r\,2}\, M_{V}^{r\,2} \, -\, F_{A}^{r\,2}\, M_{A}^{r\,2} \; = \;
v^2 \, M_{V}^{r\,2} \,\delta_{_{\rm NLO}}^{(2)}  \, ,
\end{eqnarray}
where $\delta_{_{\rm NLO}}^{(1)}$ and $\delta_{_{\rm NLO}}^{(2)}$ parameterizes the high-energy expansion of the one-loop contribution. It is then possible to fix the couplings $F_V^r$ and $F_A^r$ up to NLO.
%
\end{enumerate}

\section{Phenomenology}

\begin{figure}
\begin{center}
\includegraphics[scale=0.55]{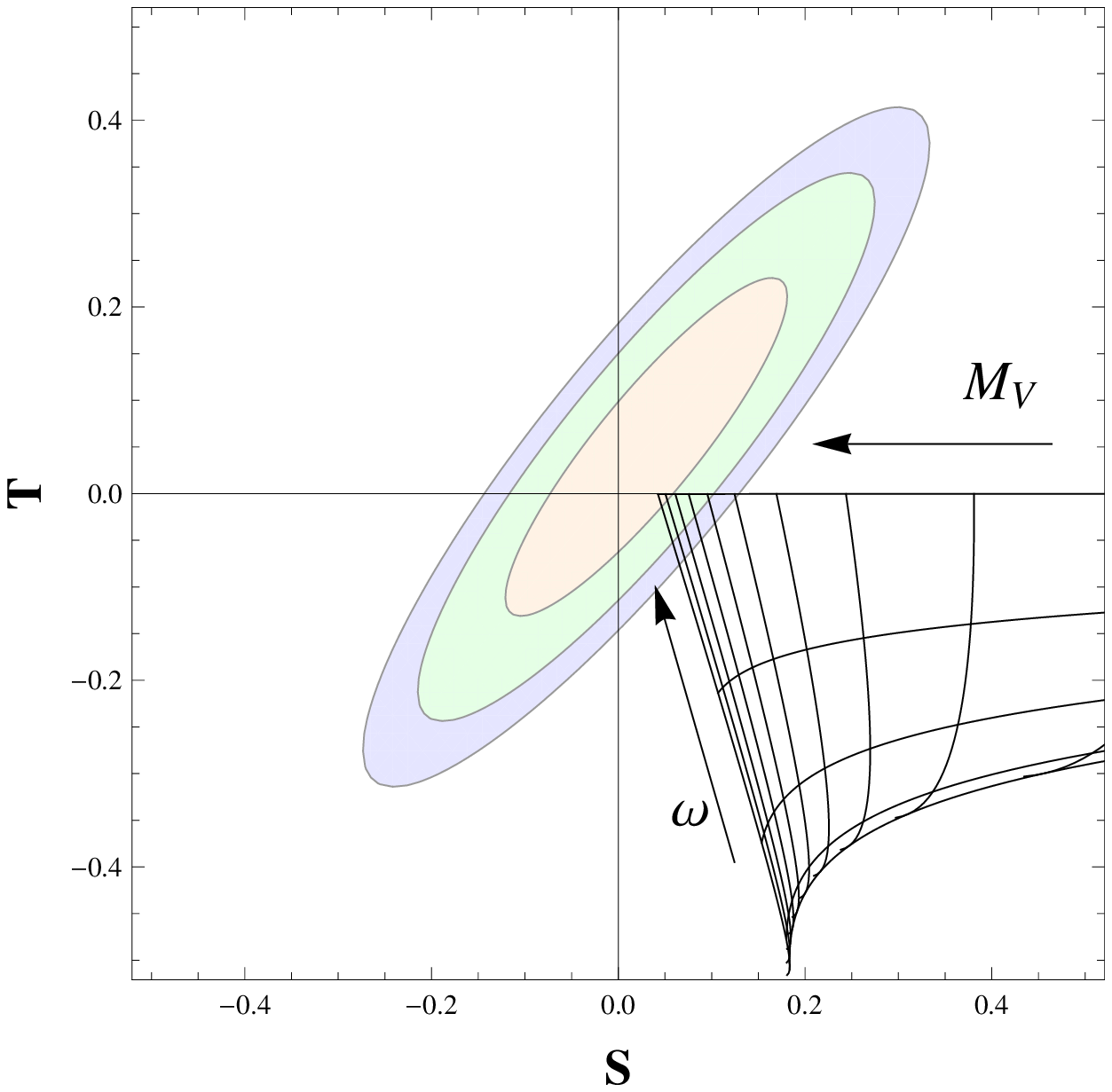} \quad \includegraphics[scale=0.60]{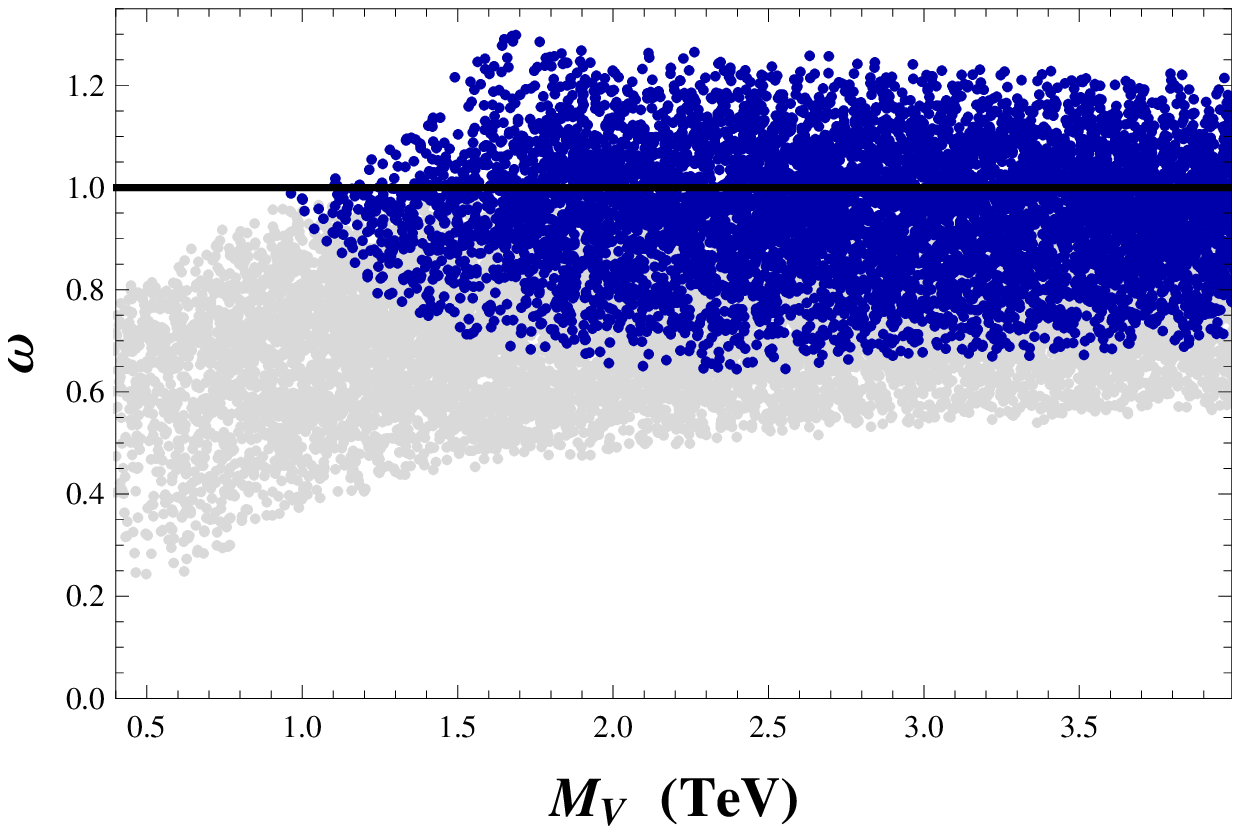}
\caption{\small{{\bf NLO determinations of $S$ and $T$, imposing the two WSRs and the VFF constraint (left)}.
The approximately vertical curves correspond to constant values
of $M_V$, from $1.5$ to $6.0$~TeV at intervals of $0.5$~TeV.
The approximately horizontal curves have constant values
of $\omega$:
$0.00, \, 0.25, 0.50, 0.75, 1.00$.
The ellipses give the experimentally allowed regions at 68\%, 95\% and 99\% CL.
{\bf Scatter plot for the 68\% CL region, in the case
when only the first WSR and the VFF constraint are assumed (right)}.
The dark blue and light gray regions
correspond, respectively,  to
$0.2<M_V/M_A<1$ and $0.02<M_V/M_A<0.2$.
}}
\label{figure}
\end{center}
\end{figure}

We have taken the SM reference point at $m_H = m_{S_1}= 126$ GeV, so
the global fit gives the results
$S = 0.03\pm 0.10$ and $T=0.05\pm0.12$, with a correlation coefficient of $0.891$~\cite{phenomenology}.
\begin{enumerate}
\item {\bf LO}. Considering the first and the second  WSRs $S_{\mathrm{LO}}$ becomes~\cite{Peskin:92}
\begin{equation}
S_{\mathrm{LO}} = \frac{4\pi v^2}{ M_V^{2}} \, \left( 1 + \frac{M_V^2}{M_A^2} \right) \,.
\end{equation}
Since the WSRs imply $M_A>M_V$, the prediction turns out to be bounded by 
$4\pi v^2/M_V^{2} < S_{\rm   LO}  <  8 \pi v^2/M_V^2$. 
If only the first WSR is considered, and assuming  $M_A>M_V$, one obtains for $S$ the lower bound 
\begin{equation}
S_{\mathrm{LO}} = 4\pi \left\{ \frac{v^2}{M_V^2}+ F_A^2 \left( \frac{1}{M_V^2} - \frac{1}{M_A^2} \right) \right\} > \frac{4\pi v^2}{M_V^2}. 
\end{equation}
The resonance masses need to be heavy enough to comply with the experimental bound.
\item {\bf NLO with the 1st and the 2nd WSRs and the VFF constraint.} With these constraints five of the seven resonance parameters are fixed and $S$ and $T$ are given in terms of $M_V$ and $M_A$~\cite{paper2}:
\begin{eqnarray}
S &=&   4 \pi v^2 \left(\frac{1}{M_{V}^2}+\frac{1}{M_{A}^2}\right) + \frac{1}{12\pi}
\bigg[ \log\frac{M_V^2}{m_{H}^2}  -\frac{11}{6}
+\;\frac{M_V^2}{M_A^2}\log\frac{M_A^2}{M_V^2}
 - \frac{M_V^4}{M_A^4}\, \bigg(\log\frac{M_A^2}{m_{S_1}^2}-\frac{11}{6}\bigg) \bigg] \,,\quad
\nonumber \\
T&=&  \frac{3}{16\pi \cos^2 \theta_W} \bigg[ 1 + \log \frac{m_{H}^2}{M_V^2}
 - \frac{M_V^2}{M_A^2} \left( 1 + \log \frac{m_{S_1}^2}{M_A^2} \right)  \bigg]  \, ,
\label{eq:T}
\end{eqnarray}
where $m_H$ is the SM reference Higgs mass adopted to define the oblique parameters and terms of $\mathcal{O}(m_{S_1}^2/M_{V,A}^2)$ have been neglected. 

In Fig.~\ref{figure} (left) we show the compatibility between the ``experimental'' values and these determinations~\cite{paper2}. The Higgs-like scalar should have a $WW$ coupling very close to the
SM one. At 68\% (95\%) CL, one gets
$\omega\in [0.97,1]$  ($[0.94,1]$),
in nice agreement with the present LHC evidence \cite{LHC}, but much more restrictive.
Moreover, the vector and axial-vector states should be very heavy (and quite degenerate);
one finds $M_V> 5$~TeV ($4$~TeV) at 68\% (95\%) CL.

\item {\bf NLO with the 1st WSR and the VFF constraint.} If only the first WSR is considered, one can still determine $T$ and obtain a lower bound of $S$ in terms of
$M_V$, $M_A$ and $\omega$~\cite{paper2}:
\begin{eqnarray}
S &\geq &  \frac{4 \pi v^2}{M_{V}^2} + \frac{1}{12\pi}  \bigg[ \log\frac{M_V^2}{m_{H}^2} -\frac{11}{6}
- \omega^2 \bigg(\!\log\frac{M_A^2}{m_{S_1}^2}-\frac{17}{6}
 + \frac{M_A^2}{M_V^2}\!\bigg) \bigg]  ,
\nonumber \\
 T&=&  \frac{3}{16\pi \cos^2 \theta_W} \bigg[ 1 + \log \frac{m_{H}^2}{M_V^2}
 - \omega^2 \left( 1 + \log \frac{m_{S_1}^2}{M_A^2} \right)  \bigg]  \, ,
\label{eq:T}
\end{eqnarray}
where $M_V<M_A$ has been assumed and again terms of $\mathcal{O}(m_{S_1}^2/M_{V,A}^2)$ have been neglected.

Fig.~\ref{figure} (right) gives the allowed 68\% CL region in the space of parameters $M_V$ and $\omega$,
varying $M_V/M_A$ between 0 and 1~\cite{paper2}. Note, however, that values of $\omega$
very different from the SM can only be obtained with a large
splitting of the vector and axial-vector masses. In general there is no solution for $\omega >1.3$.
Requiring $0.5<M_V/M_A<1$, leads to $1-\omega <0.16$
at 68\% CL, while the allowed vector mass stays above $1.5$~TeV.
\end{enumerate}

In summary, strongly-coupled electroweak models with massive resonance states
are still allowed by the current experimental data.
Nonetheless, 
the recently discovered Higgs-like boson
with mass $m_{S_1}=126$~GeV must have a $WW$ coupling close to the SM one ($\omega=1$).
In those scenarios, such as asymptotically-free theories,
where the second WSR is satisfied, the $S$ and $T$ constraints force $\omega$ to be in the range
$\left [ 0.94, 1\right]$ at 95\% CL. Larger departures from the SM value can be accommodated when
the second WSR does not apply, but one needs to introduce a correspondingly large mass splitting between the vector and axial-vector states.

\vspace{0.5cm}


\end{document}